\documentclass[10pt,english,conference]{IEEEtran}

\usepackage[dvipsnames]{xcolor}
\usepackage{etex}
\usepackage{pstricks}

\usepackage{adjustbox}
\usepackage[T1]{fontenc}
\usepackage[latin9]{inputenc}
\usepackage{listings}
\usepackage{float}
\usepackage{amssymb}
\usepackage{graphicx}
\usepackage{url}
\usepackage{tabularx}
\usepackage{amsmath}
\usepackage{babel}
\usepackage[hidelinks]{hyperref}
\usepackage{algorithm}
\usepackage{tikz}
\usepackage{epstopdf}
\usepackage{tkz-graph}
\usepackage{amsmath}
\usepackage{filecontents}
\usepackage{flushend}

\usetikzlibrary{external}
\definecolor{arsenic}{rgb}{0, 0, 0.6}
\definecolor{findingOnly}{gray}{0.4}
\definecolor{findingBoth}{gray}{0.8}
\definecolor[named]{Back}{cmyk}{0,0,0,.3}
\definecolor[named]{Front}{cmyk}{1,1,1,.34}
\definecolor{javared}{rgb}{0.6,0,0} 
\definecolor{javagreen}{rgb}{0.25,0.5,0.35} 
\definecolor{javapurple}{rgb}{0.5,0,0.35} 
\definecolor{javadocblue}{rgb}{0.25,0.35,0.75} 

\usetikzlibrary{arrows,automata}
\usetikzlibrary{shadows}
\usetikzlibrary{positioning}

\makeatletter
\pgfdeclareshape{document}{
\inheritsavedanchors[from=rectangle] 
\inheritanchorborder[from=rectangle]
\inheritanchor[from=rectangle]{center}
\inheritanchor[from=rectangle]{north}
\inheritanchor[from=rectangle]{south}
\inheritanchor[from=rectangle]{west}
\inheritanchor[from=rectangle]{east}
\backgroundpath{
\southwest \pgf@xa=\pgf@x \pgf@ya=\pgf@y
\northeast \pgf@xb=\pgf@x \pgf@yb=\pgf@y
\pgf@xc=\pgf@xb \advance\pgf@xc by-12.5pt 
\pgf@yc=\pgf@yb \advance\pgf@yc by-12.5pt
\pgfpathmoveto{\pgfpoint{\pgf@xa}{\pgf@ya}}
\pgfpathlineto{\pgfpoint{\pgf@xa}{\pgf@yb}}
\pgfpathlineto{\pgfpoint{\pgf@xc}{\pgf@yb}}
\pgfpathlineto{\pgfpoint{\pgf@xb}{\pgf@yc}}
\pgfpathlineto{\pgfpoint{\pgf@xb}{\pgf@ya}}
\pgfpathclose
\pgfpathmoveto{\pgfpoint{\pgf@xc}{\pgf@yb}}
\pgfpathlineto{\pgfpoint{\pgf@xc}{\pgf@yc}}
\pgfpathlineto{\pgfpoint{\pgf@xb}{\pgf@yc}}
\pgfpathlineto{\pgfpoint{\pgf@xc}{\pgf@yc}}
}
}
\makeatother
\tikzset{
    state/.style={
           rectangle,
           rounded corners,
           draw=black, very thick,
           minimum height=2em,
           inner sep=5pt,
           text centered,
           },
    message/.style={
           document,
			fill=white,
			line width=1pt,
           inner sep=5pt,
           draw=black, very thick,
           minimum height=2em,
           inner sep=2pt,
           text centered,
           },
}

\usepackage{amsmath, listings, amsthm, amssymb, proof, xspace}

\newcommand{\alt}{~~|~~}

\newcommand{\infr} [3] [] {\infer[\textsc{#1}]{#3}{#2}}

\mathchardef\mhyphen="2D

\begin{document}

\title{LUCON: Data Flow Control for Message-Based IoT Systems}
\author{
    \IEEEauthorblockN{Julian Sch\"{u}tte, Gerd Stefan Brost}
    \IEEEauthorblockA{Fraunhofer AISEC, Germany
    \\\{julian.schuette,gerd.brost\}@aisec.fraunhofer.de}
}
\maketitle

\begin{abstract}
Today's emerging Industrial Internet of Things (IIoT) scenarios are characterized by the exchange of data between services across enterprises.
Traditional access and usage control mechanisms are only able to determine \emph{if} data may be used by a subject, but lack an understanding of \emph{how} it may be used. The ability to control the way how data is processed is however crucial for enterprises to guarantee (and provide evidence of) compliant processing of critical data, as well as for users who need to control if their private data may be analyzed or linked with additional information -- a major concern in IoT applications processing personal information.
In this paper, we introduce \mbox{LUCON}, a data-centric security policy framework for distributed systems that considers data flows by controlling how messages may be routed across services and how they are combined and processed.
LUCON policies prevent information leaks, bind data usage to obligations, and enforce data flows across services. Policy enforcement is based on a dynamic taint analysis at runtime and an upfront static verification of message routes against policies. We discuss the semantics of these two complementing enforcement models and illustrate how LUCON policies are compiled from a simple policy language into a first-order logic representation. We demonstrate the practical application of LUCON in a real-world IoT middleware and discuss its integration into Apache Camel. Finally, we evaluate the runtime impact of LUCON and discuss performance and scalability aspects.
\end{abstract}

\section{Introduction}
\label{sec:introduction}
While IoT systems in general create undeniable benefits in areas like health care, home automation, manufacturing, logistics, and mobility, it is also obvious that existing threats to the integrity of business processes and the privacy of users intensify with an increasing degree of distribution, amount of endpoints and trust domains. A paramount challenge for data owners is to control the way how their data is processed and combined with data from other sources, and how it is published to untrusted third parties.

Today, Industrial IoT systems are characterized by data flowing from sensors to services and applications and possibly back to actuating devices. These data flows span several physical platforms, including resource-constrained sensors, mobile devices, and cloud backends. In contrast to traditional enterprise systems, modern distributed IoT systems typically span several ''trust domains'', i.e. within a single application, data is processed by services under different authoritative controls.

In addition, privacy and security are interleaved since sensor data may contain personal data of employees (e.g., a machine operator) or private users (e.g., the owner of a home automation solution).

Traditional access control cannot cope with this challenge -- it merely aims at controlling actions of subjects to resources (e.g., a ''read'' request from a user to a file). In that sense, traditional access control is \emph{resource-centric} and unaware of the actual processing of data, as access to resources is only controlled at a specific point in time without further control on how it is used. Further, typical access control languages like XACML provide means to describe resources, but do not allow to write rules referring to classes of data that provided by these resources. It is impossible to state that only certain information may be retrieved from an endpoint, while some other information must not be published by the same endpoint.

An extension of access control is \emph{usage control} which has been introduced in the early 2000s \cite{Park2004} and has been subject to intensive research in the following decade \cite{hilty2007,katt2008,Lazouski2010}. Usage control extends access control by the dimension of time and is able to continuously monitor and control the usage of resources such as files or services by subjects. However, it is mostly still resource-centric as it only decides access requests to resources in the course of time, but does not control how sensitive information is processed and combined. ABAC languages like XACML 3.0 are moving in that direction by supporting the notion of \emph{obligations} which must be fulfilled by the subject. The outcome of an obligation does however not influence the policy decision anymore and is out of the semantics of XACML. 

\begin{figure}[tb]
  \centering
  \includegraphics[width=.8\columnwidth]{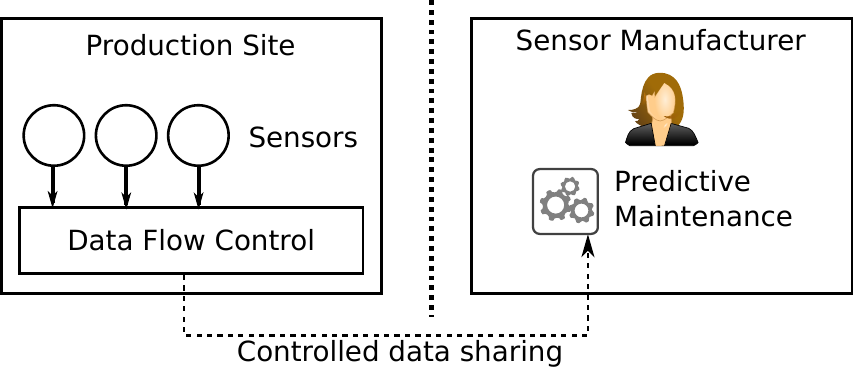}
  \caption{Data flow control for predictive maintenance}
  \label{fig:intro-scenario}
\end{figure}

In modern data-centric systems, the concept of resource-centric protection does not apply anymore. As a consequence, traditional usage control models fall short of enforcing requirements of data owners. With a growing number of data sources in the form of sensors from different owners and cloud-based data analytics services, the challenge is not anymore to control the usage of a single resource, but to express constraints on how \emph{data objects} (messages) may be processed .
\autoref{fig:intro-scenario} illustrates a typical scenario: sensors in a production facility measure parameters of the production process like flow rate and temperature of various liquids. These measurements are essential for controlling the production process, but they are also interesting for analytics applications from third parties. Knowing these measurements helps manufacturers developing ''predictive maintenance services'' such as the detection of sensor drifts or an approaching end of life of their hardware.. However, up to date, these scenarios are hardly possible due to the sensitive nature of raw sensor data, from which trade secrets like recipes, production processes and capacities can immediately be derived.

So, controlling access to resources (in this case, sensors) or their data is not sufficient. Rather, it is necessary to control the flow of data, i.e. the way how it is combined, processed and shared with different endpoints. In the example from \autoref{fig:intro-scenario}, it must be guaranteed that sensor manufacturers only get data from specific sensors. This data must pre-processed in a way that allows them to run their analytics, but not to reverse-engineer the production process or any other (including privacy related) data.

In this paper we introduce LUCON, a policy framework for data flow control (DFC) in distributed systems, which provides a runtime monitor for dynamic enforcement of DFC policies and an upfront static verification of message routes against policies. As discussed in \cite{schuette2016}, we model all data exchanges as message flows. Data flow policies are based on a formal system model and an operational semantics of the enforcement in message routes, which pre-configure possible data flows. After introducing the formal foundation of LUCON we show how it is implemented in a real-world messaging system and supports both static and dynamic enforcement. The benefit of an upfront static enforcement is that users can analyze potential, possibly counter-intuitive violations of their policy, while in a dynamic enforcement only concrete violations of policies will be prevented. In our policy framework, the result of a policy enforcement is either a simple cancellation of a message flow or the execution of an obligation, i.e. an action which must be taken by the enforcement component to fulfill the policy. By means of a prototype implementation of the LUCON framework and its integration into the Apache Camel messaging router, we show that the performance of LUCON is suited for large-scale productive applications.

\section{Related Work}

Usage control has been subject to extensive research for quite a while \cite{sandhu2003usage}. While several models have been proposed, the most prominent one is $UCON_{ABC}$, originally introduced by Park and Sandhu \cite{Park2004}. It comprises Authorizations (A), oBligations (B), and Conditions (C), referring to attributes of subjects and resources. Attributes are mutable (e.g., they can change over time) and the continuity of access decisions is formalized. In this way,  UCON A, B and C can be defined to be evaluated before (pre) or during usage (on). The model has undergone different extensions in the course of time. As an example, \cite{katt2008} incorporated post-obligations.  $UCON_{ABC}$ does not dictate how to design a specific architecture and mechanisms for usage control, but stays abstract in that manner. Other approaches focus on specific languages rather than abstract models, such as the Obligation Specification Language (OSL) \cite{hilty2007}.

Much has been done in the area of formalization of usage control policies and the formal analysis of their properties.
In \cite{Zhang2005}, a formalization of $UCON_{ABC}$ in Lamport's Temporal Logic of Actions (TLA) is given, in \cite{Basin2010,Basin2012} Basin et al. give an approach on analyzing usage control policies formalized in first-order temporal logic (MFOTL). In \cite{Pretschner2009}, a Linear Time Logic (LTL) dialect is used for the sake of analyzing policies, and in \cite{Elrakaiby2014} an analysis of dynamically changing usage control policies is described, based on Action Computation Tree Logic (ACTL). Our work is based on this research, but the approach is more specific and focuses on the application of usage control to data processing only.\\

The concept of enforcing data flow control in decentralized systems has already been introduced by Myers et al.~\cite{Myers1997}. Their understanding of controlling information flows refers to preserving secrecy and integrity properties of classified documents -- an approach that pursues the enforcement of traditional information classification systems such as the Bell LaPadula model (\emph{no read up, no write down}) for secrecy and the Biba model \cite{Biba1977} (\emph{no read down, no write up}) for integrity. Myers proposes a label-based approach to mark data sets and to prevent information leakage by annotating existing programming languages. We generalize this concept by concentrating on information flow between components (services), that have no built in mechanisms for supporting external labels.
The enforcement of usage control policies is a central challenge, as it requires system-specific implementations and trust relationships related components. Trustworthy system architectures for usage control enforcement have been proposed in \cite{Zhang2008}, which allow usage control policies at the level of system calls, given that the trustworthiness of the enforcement point can be attested using hardware-based mechanisms. Our approach does not focus at usage control enforcement on remote platforms, but rather on a specific enforcement mechanism for data processing. Nevertheless, it could be combined with techniques from \cite{pretschner2006distributed} or \cite{Zhang2008} in case the enforcement would have to take place on remote hosts.\\
Closer related to our work is  \cite{Harvan2009}, which introduces the idea of using data flow tracing at the level of system calls in order to enforce usage control policies. The authors show that based on an underlying data flow model, more realistic and expressive policy rules can be written, referring to states of a data flow system, rather than specific sequences of events.
In \cite{Pretschner2009a}, this approach is extended by tracing messages in the X11 environment, specifically copy \& paste actions on sensitive data which is either blocked or replaced by meaningless data in case a policy is violated. Similar to \cite{Harvan2009,Pretschner2009}, we understand usage control as enforcing conditions in data flows.

Fine granular data flow tracking for databases has been done by applying taint tracking in \cite{Davis2010} for specific applications. A similar approach was followed by \cite{Chinis2013}, providing an API that allows to introduce taint tracking for legacy web applications without major code changes. This is also based on taint tracking at database level and uses hooks that are placed in legacy code to enable security policy enforcement.

Pasquier et al. proposed CamFlow \cite{Pasquier2015}, an end-to-end information flow control enforcement system for cloud systems based on the implementation of Linux Security Modules as enforcement points. Thus, CamFlow is tightly integrated into the operating system via a custom Linux Security Module and considers information flows between processes in an OS. This work has been continued in \cite{Pasquier2016} with a focus on DETA (Declassify, Endorse, Transform, Authorize) policies.
Apart from the fact that CamFlow is an operating-system-level mechanism while our approach mainly addresses message buses in a distributed system, CamFlow proposes fixed security rules for secrecy and integrity, following traditional information classification concepts. Our system, in contrast, proposes a more generic labeling mechanism that allows to create any information class, track its usage in the system and write policies on how to handle read and write accesses to it.


\section{Overview of our approach}
\label{sec:overview}

LUCON is a policy framework to enforce secure data flows in message-based systems -- typically in IoT architectures. Its design goals are a reasonable low runtime overhead, a formal semantics and support of authors in writing flawless policies for existing message routes. It is comprised of the following components:

\begin{itemize}
  \item the definition of a policy language, its implementation in Eclipse XText, and its compilation into a first-order logic representation
  \item a runtime evaluation of policies based on the first-order logic representation of policies, following a message tainting approach with a formal execution semantics of policy-controlled message routes
  \item a static model checking of message routes against policies in the first-order logic representation, including a compilation of Apache Camel message routes into Prolog programs
\end{itemize}

\subsection{Policy language} The motivation for a policy language is to separate the specification of security requirements on data flows from the actual messaging system. Existing DFC systems like \cite{Pasquier2015,Jif2001,Pasquier2014} enforce predefined flows between security classes. However, in practice users have diverse and application-specific requirements which cannot be hard coded into a generic distributed middleware. Early research on information flow control has proposed various information classification models such as Chinese Wall, Biba or Bell LaPadula, where each serves one specific requirement (such as either integrity or secrecy for lattice-based information classifications), but they are not necessarily compatible with one another. Thus, instead of hard coding valid flows into the system, we rather aim for a simple domain specific language (DSL) to allow the user to define custom policies. The DSL compiles into a logic representation in Prolog -- a programming language  based on Horn clauses which is a decidable subset of first-order logic. The benefit of that is that compiled policies have a formal foundation and can be used to model-check requirements against message routes, but at the same time can be efficiently evaluated at runtime so that performance impact on a productive system remains low.

\subsection{Runtime enforcement} Runtime enforcement implements a dynamic taint-style analysis and thus leans towards the permissive end of possible data flow enforcement strategies, as we will discuss in \autoref{sec:hybrid-information-flow-control}.
The aim of LUCON is to prevent messages from violating the policy, but not to prevent any information leaks over side channels. Some data flow systems proposed in the past \cite{Sabelfeld2010} apply a stricter strategy and block even information leaks over side-channels, for example in cases in which the attacker can learn private information by observing the control flow or exceptional terminations of a message route. However, these approaches assume that the attacker knows the exact control flow specification (i.e., the message routes), which is not the case in our model. Second, implicit information leaks are not intuitive for the user and sudden cancellation of a route at runtime may not be expected behavior. In our taint-style approach we mark messages with an initial set of labels as soon as they are created. Labels are transported along with messages and possibly altered when the message is processed by a service. The mechanism for the modification of message labels is called the \emph{taint propagation logic} and determines how the security and privacy properties of a message change as it is processed and merged by services. In our approach, the definition of the taint propagation logic is part of the policy. This allows our runtime monitor to query the compiled policy for the changes which should be made to message labels, as well as for the actual data flow policy.

\subsection{Static model-checking} Runtime enforcement is tuned to be fast and will prevent messages from leaking information by blocking or modifying them just before the data leak would occur. For users, it is however important to analyze if and under which circumstances their message system would run into potential data leaks so they can verify that message routes will not be unexpectedly terminated by the policy framework. Further, LUCON will provide evidence that message routes fulfill the security requirements, which is important information for audit and compliance purposes. This is achieved by translating message routes into a first-order logic representation, analog to the representation of compiled policies. The logic model allows to verify route definitions against policies so that users can check upfront whether routes are applicable at all under a certain policy, whether only specific executions paths may violate the policy, or whether a route is fully compliant with a policy. In case of potential policy violations, LUCON will generate an example of a message flow violating the policy.


\section{System Model}
\label{sec:system-model}

We first set the common ground for the abstract type of system that is addressed by our policy framework. In practice, LUCON runs in any message-based IoT system, but for the remainder of this paper we establish an understanding of the terms and concepts that are relevant in those systems.

The system is based on \emph{services} which communicate via \emph{messages}. A service accepts a set of input messages, operates on their content and emits a set of output messages. Each service is under control of a trust domain and as messages are sent from one service to another, they may cross domain boundaries. The ability of a user to define and apply policies is limited to their own trust domain, i.e. policies of a user can only control messages within their domain. With respect to enforcement of policies, however, it is still possible that a user retrieves an assertion of a successful enforcement from a remote domain -- either by establishing trust at a technical level (e.g., by remote attestation) or by retrieving evidence of the enforcement (e.g. by observing expected side-effects).

Both, We denote these sets of predicates as message \emph{labels} $\mathcal{L}$ and service \emph{properties} $\mathcal{P}$, respectively.

\paragraph{Message Labels}
Message labels $\mathcal{L}$ classify a message in terms of its data source or secrecy level and may be partially ordered. For instance, a message $m$ can be labeled as $\mathcal{L}_m = \{\texttt{classification(top\_secret)}\}$  (1-ary predicate) or $\mathcal{L}_m = \{\texttt{personal\_data}\}$ (0-ary predicate). The specific predicates are not determined by the model but rather by its instantiation in a specific application.

\paragraph{Service Properties}
Service properties $\mathcal{P}$ are used to describe services. For example, a service which stores data in a database can be assigned the predicates $\mathcal{P} = \{\texttt{persist}\}$ (0-ary predicate) or $\mathcal{P} = \{\texttt{persist({\footnotesize jdbc://localhost/...)}}\}$ (1-ary predicate).


\paragraph{Message Routes}
The interaction between services is defined as an Enterprise Integration Pattern (EIP) \cite{Hohpe2003} in form of a message \emph{route}. We consider a route as a non-while-looping program, i.e. a sequence of numbered statements which either call external services, assign values to variables, or control execution of the next statement. Note that excluding while loops from our route definition is a limitation compared to the expressiveness of real-world turing-complete message routers like Apache Camel or Spring Integrations, which do in fact allow the construction of while loops in EIPs like \emph{Dynamic Router}\footnote{\url{http://www.enterpriseintegrationpatterns.com/patterns/messaging/DynamicRouter.html}}. The reason we chose to exclude while-loops is that it turns the static route verification into a decidable problem, while in practice while-loops are rarely used in EIPs and e.g. discouraged by Apache Camel\footnote{cf. \url{http://camel.apache.org/loop.html}}. 
Routes support variables in two scopes: global and message-scoped. Global variables are available across all executions of a route, while message-scoped variables get appended to the message object and are transported along with it. Branching statements refer to conditions over variables and fork the control flow into several branches, just like conditions in a program.

\begin {figure}
       \centering
       \resizebox {\columnwidth} {!} {
       \begin{tikzpicture}[->,shorten >=1pt,auto,node distance=2.9cm,thick,main node/.style={inner sep=8pt,rounded corners=.1cm,draw,align=center}, main edge/.style={}]
\usetikzlibrary{shapes.misc}
\tikzset{
   -|/.style={to path={-| (\tikztotarget)}},
   |-/.style={to path={|- (\tikztotarget)}},
}
 \node[main node] (source) [] {\textbf{Sensor}\\(\textit{from})};
 \node[main node] (split) [right of=source] {\textit{split}};
 \node[main node] (proc) [right of=split] {\textbf{Log}\\(\textit{to})};
 \node[main node] (anon) [above of=proc,node distance=2.5cm] {\textbf{Merge}\\(\textit{bean})};
 \node[main node] (aggr) [right of=proc] {\textit{aggregate}};
 \node[main node] (sink) [right of=aggr] {\textbf{Outbound Queue}\\(\textit{to})};

 \path
   (source) edge node [below]{} (split)
   (split) edge [|-] node [below]{} (anon)
    edge node [below]{} (proc)
   (proc) edge node[right] {} (aggr)
   (anon) edge [-|] node[right] {} (aggr)
   (aggr) edge node [right]{} (sink);
\end{tikzpicture}}
\caption{A message route over several services}
\label{fig:example-route}
\end{figure}
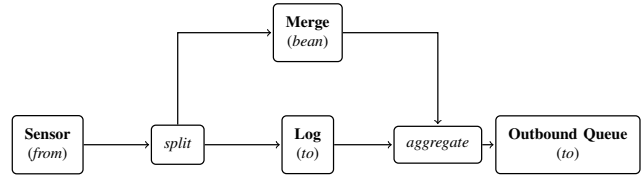

Accordingly, the set of supported statements comprises variable assignments (\texttt{set-*-prop}), control flow modification (\texttt{choice}), message manipulation (\texttt{split,aggregate,bean}), and service invocation (\texttt{from, to}). The simplified grammar is given in the following listing and \autoref{fig:example-route} depicts an example route.

{\footnotesize
 \[
   \begin{array}{rcl}
   stmt    &:=& \emph{assign-msg} \alt \emph{assign-env} \alt \emph{from} \\
           &  & \alt \emph{to} \alt \emph{choice} \alt \emph{split} \alt \emph{aggregate}\\
   assign\mhyphen msg  &:=& \texttt{set-msg-prop}~\text{var} := \emph{expr}\\
   assign\mhyphen env  &:=& \texttt{set-env-prop}~\text{var} := \emph{expr}\\
   from    &:=& \texttt{from(}\emph{service}\texttt{)} \\
   to      &:=& \texttt{to(}\emph{service}\texttt{)} \\
   choice  &:=& \texttt{when}~\emph{expr}~\texttt{then goto}~\emph{v}\\
           &  & \texttt{otherwise goto}~\emph{v}\\
   split   &:=& \texttt{split}~\emph{expr}\\
   aggregate &:=& \texttt{aggregate}~\emph{expr}\\
   \emph{expr} &:=& \text{n-ary Prolog predicate}\\
   \emph{v} &:=& \text{Statement number}\\
   \emph{service} &:=& \text{Service name}\\
   \end{array}
 \]
}

To model execution of a route, we further introduce some \emph{execution contexts} which represent the current state of the execution: $\Sigma$ maps statement numbers to statements. $\mu_m$ holds the message-scoped variables and maps each variable of a message $m$ to its value. A global map $\Delta$ assigns global variable names to their values. Further, a program counter \texttt{pc} holds the number of the currently executed statement and an instruction pointer $\iota$ the number of the next statement.

\autoref{tab:execution-contexts} summarizes these execution contexts.

\begin{table}[tb]
  \caption{Execution contexts}
  \label{tab:execution-contexts}
  \centering
  \begin{tabular}{|ll|}
    \hline
    $\tau$      & Maps a message to its taint state, e.g. $\tau[m \leftarrow 1] \Rightarrow 1=\tau[m]$      \\
    $\Sigma$    & Maps a statement number to a statement      \\
    $\mu_m$     & Maps variables of message $m$ to their value\\
    $\Delta$    & Maps global variable names to their current value\\
    \texttt{pc} & Number of the currently executed statement\\
    $\iota$     & Number of the next statement \\
    \hline
  \end{tabular}
\end{table}

\paragraph{Trust Domains}
Services and routes reside in \emph{trust domains}. A trust domain is controlled by a single authority that can create and update route definitions and it is assumed that services within a trust domain behave correctly in terms of propagating message labels. It is important to note that we assume that route definitions are not known outside of the trust domain. If this assumption would not hold, information from messages may leak if an attacker is able to observe the control flow, i.e. the execution of a route.


\section{Hybrid Information Flow Control}
\label{sec:hybrid-information-flow-control}

LUCON takes a hybrid approach on information flow control by combining a dynamic and a static component: The dynamic policy enforcement is tuned for efficiency and to limit  delays by policy evaluations at runtime. It is based on a \emph{taint-style analysis} that prevents explicit information leaks under the assumption that implicit leaks (for example by control flow observation or untrusted misbehaving services) do not occur. A static, upfront model-checking verifies message routes against policies and informs the user about potential policy violations. Here, runtime performance does not play a role, but rather completeness of the verification and the generation of understandable counterexamples so as to either guarantee that message routes are free of policy violations (e.g., for audit purposes) or to support policy authors in fixing potential flaws.

\subsection{Static vs. Dynamic Data Flow Control}

Data flow research dates back to the seventies when Denning \cite{Denning74} proposed a lattice-based organization of security classifications to mathematically formulate constraints on information flows -- a formal foundation for the Bell-LaPadula secrecy \cite{BellLaPadula73} and Biba \cite{Biba1977} integrity models. Since then, various data flow control systems have been proposed, either relying on static checking of system configurations against information flow rules, or on dynamic enforcement of data flow constraints at runtime \cite{Pasquier2014,Pasquier2015,Jif2001,FlowCaml2003}. These systems typically enforce secrecy by preventing information leaks over explicit flows and partly over implicit flows. Explicit flows refer to leakage of information directly into publicly readable sinks, whereas implicit leaks refer to side-channel leaks via control flow or termination. Sabelfeld et al. show in \cite{Sabelfeld2010} that dynamic enforcement and the classic Denning-style static flow control can only achieve \emph{termination-insensitive non-interference}, i.e. they can prevent information leaks via observation of the control flow by an attacker, but not via observation of route termination. Taint analysis, as we adopt for our dynamic runtime enforcement, provides even weaker guarantees as it allows control-flow leaks in some cases. To illustrate this, let us consider the following route, which we denote as a simple program, for the sake of readability.

\begin{lstlisting}[xleftmargin=2em]
tainted := ...;  // Taint label set
public := 1;
tmp := 0;
if tainted then
  tmp := 1;
if tmp != 1 then
  public := 0
\end{lstlisting}

In this example, sensitive information is written into a variable \texttt{tainted}, which will consequently be marked with a taint label (line 1). As the value of \texttt{tainted} is never assigned to any other variable, the taint flag is not propagated. Variable \texttt{public} is written into a public data sink and thus leaks its content. As can be seen from the example, the value of variable \texttt{public} is equal to \texttt{tainted} in all possible execution paths, although it is never explicitly assigned. Consequently the route leaks tainted information to a public data sink and a classic taint analysis is not able to detect this leak.

A Denning-style type system would behave differently and manage a stack of global security contexts. In line 4, when a ''secure'' condition is evaluated, a new \emph{secure} item would be added to the stack. When execution enters line 5, the system would notice that a non-secure variable is written within a secure context and would terminate execution immediately, thereby preventing the information leak. Denning-style type systems are thus more strict and prevent information leaks even under the assumption that the attacker know the control flow specification (the message route) and is able to observe control flow at runtime. However, in the type of distributed system we address, Denning-style flow control is less appropriate, as it introduces a variety of practical issues:
first, the assumptions are overly strict. Attackers might be able to observe parts of the control flow, e.g. by hosting a service which is used by a message route, but they do not know the control flow specification (the message routes) nor can they learn it by globally observing the control flows of domain. Second, in real-life message routes, all operations would have to be considered as write operations, as they are typically realized by some implementation which is not further known to the data flow control engine. If all operations are writes, however, a single access to a non-tainted variable will lead to termination of the route. This is unexpected for the user, at best, and will render the system dysfunctional.

As a consequence, LUCON adopts a dynamic taint approach which is efficient at runtime and prevents explicit leaks of information, i.e. it prevents routes from processing data in an undesired way, assuming that an attacker cannot retrieve the route specification and globally observe the control flow. Complementary to runtime enforcement, LUCON provides an upfront static model checking to verify message routes against policies. In the approach we describe herein, the model checking follows the same taint-style semantics as the dynamic enforcement, but in general the models do not have to be equal. For instance, it would be possible to statically verify routes in a stricter control-flow- and termination-sensitive model to identify even theoretical information leaks, while still running dynamic enforcement in the more realistic and relaxed taint-style model.

\subsection{Dynamic Taint-Style Enforcement}

The basic idea of the taint-style dynamic flow enforcement is to assign a set of taint labels to messages when they enter the system and to modify the taint labels as messages are processed by services. Whenever a message is about to be sent to an external service, the policy is consulted to check whether a respectively marked message may enter this specific service.

Different from other information and data flow control systems, LUCON does not dictate a set or lattice of taint labels, but rather allows to assign any set of labels to messages. Taint labels are represented as first-order logic predicates and any rule over these predicates can be declared to construct lattices, hierarchies or any other inference of labels. Assignment of labels to messages is controlled by the taint propagation logic, which is part of the policy. In fact, every service description in the policy may include two \emph{label transformation functions} $\mathcal{L}^-(\cdot)$ and $\mathcal{L}^+(\cdot)$ that determine which labels will be removed and added to a message, respectively. To denote the semantics of a route with taint tracking enabled, we introduce an additional context $\tau$ that maps variables to the set of taint labels assigned to them. That is, $\tau_\Delta$ denotes the taint states of global variables and $\tau_m$ denotes the taint labels assigned to a message $m$. Individual variables of a message cannot be tainted, rather the whole message will be marked.

\begin{figure*}
\label{fig:tainted-msg}
\caption{Message $m$ with taint labels $L_m$ sent along services with properties $P$}
\vspace*{.3cm}
\begin{center}
\begin{adjustbox}{width=.85\linewidth}
\begin{tikzpicture}[->,>=stealth']

 \node[message] (MESSAGE)
 {\begin{tabular}{l}
  \textbf{Message $m$}\\
  ~\\
  \parbox{5.5cm}{
  $L_m=\{raw,temperature\}$\\
  $\textit{Policy}=\{publish \leftarrow \neg raw\}$\\
  }
 \end{tabular}};

 \node[state,     
  yshift=0cm,     
  below of=MESSAGE,   
  node distance=4cm,  
  anchor=center] (SERVICE_A)  
 {%
 \begin{tabular}{l}   
  \textbf{Service A}\\
  (Database)\\
  ~\\
  \parbox{5.5cm}{
    \footnotesize
    $\mathcal{L}^- = \emptyset$\\
    $\mathcal{L}^+ = \emptyset$\\
    $L_m=\{raw,temperature\}$\\
    $P = \{persist(\texttt{hdfs2://...})\}$

  }
 \end{tabular}
 };

 \node[state,
  right of=SERVICE_A,
  node distance=7cm,
  anchor=center] (SERVICE_B)
 {%
 \begin{tabular}{l}
  \textbf{Service B}\\
  (Merges Data)\\
  ~\\
  \parbox{4.8cm}{
    \footnotesize
    $\mathcal{L}^-=\{raw\}$\\
    $\mathcal{L}^+=\{merge(10)\}$\\
    $L_m=\{temperature,merge(10)\}$\\
    $P = \emptyset$
  }
 \end{tabular}
 };

 \node[state,
  right of=SERVICE_B,
  node distance=7cm,
  anchor=center] (SERVICE_C)
 {%
 \begin{tabular}{l}
  \textbf{Service C}\\
    (Publisher)\\
  ~\\
    \parbox{5.5cm}{
      \footnotesize
      $\mathcal{L}^-=\emptyset$\\
      $\mathcal{L}^+=\emptyset$\\
      $L_m=\{temperature,merge(10)\}$\\
      $P = \{publish(\texttt{http://...})\}$
    }
 \end{tabular}
 };

 \path (MESSAGE)    edge  node[anchor=west,left]{m'}   (SERVICE_A)
       (SERVICE_A)  edge  node[anchor=south]{m'}       (SERVICE_B)
       (SERVICE_B)  edge  node[anchor=south]{m'}       (SERVICE_C)
;

\end{tikzpicture}
\end{adjustbox}
\end{center}
\end{figure*}
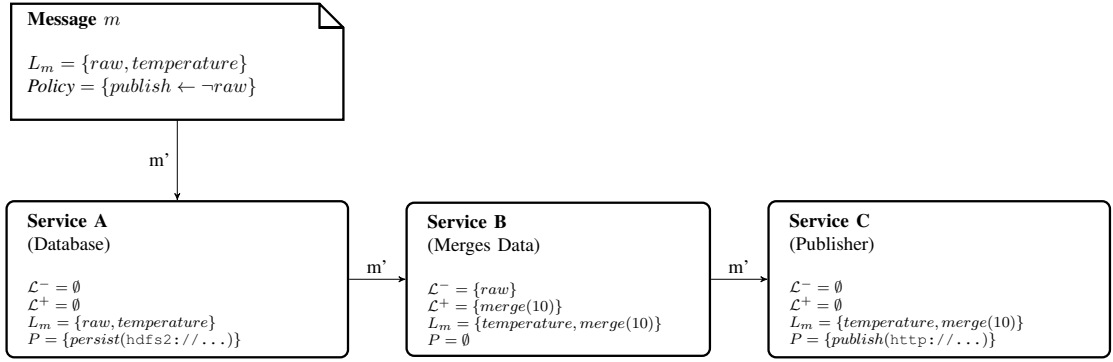

The operational semantics of a taint-controlled route execution is given in the appendix in \autoref{fig:operational-semantics}. Inference rules are written as
~\\
\begin{center}
$\infr{Computation}{\langle Current~state \rangle, Stmt \to \langle Next~state \rangle}$\\
\end{center}
~\\
where $Current~state$ and $Next~state$ are written as tuples $\langle \tau, \Sigma, \mu_m, \Delta, \texttt{pc}, \iota\rangle$ and denote the system state before and after execution of the statement $stmt$, respectively. $Computation$ denotes the actual computation on the system state which is applied by executing $Stmt$. The notation of computations makes use of expressions in the form $\mu_m, \Delta \vdash e \Downarrow v$, which means that an expression $e$ evaluates to value $v$ in the context denoted by messages properties $\mu_m$ and system variables $\Delta$.

Statements refer to operations of typical enterprise integration patterns (EIP), as used by Apache Camel\footnote{\url{http://camel.apache.org/schema/spring/camel-spring-2.19.2.xsd}}. The formal semantics covers far from all Camel EIPs but is focused on the statements relevant for information flows.

 The \textsc{From} statement reads data from a service endpoint and \textsc{To} and \textsc{bean} forward it to an external service or an internal processing bean, respectively (a component that may modify the message). \textsc{Choice} denotes a branch in the control flow and is similar to an if-then-else-statement in a normal program. With \textsc{Split}, a message can be split by an expression into multiple messages which are processed in parallel and can be joined again by \textsc{Aggregate}. Statements \textsc{assign-msg-prop} and \textsc{assign-env-prop} set message-scoped and global-scoped variables to the value of a given expression.  Variables are only visible within the message routing engine and not delivered to actual services, therefore they do not affect the taint state of a message. Rather, the only statements affecting the taint state $\tau$ are \textsc{From}, \textsc{To}, \textsc{Bean}, \textsc{Split}, and \textsc{Aggregate}. When a message is created by \textsc{From}, it is assigned the taint labels determined by the taint policy $\mathcal{L}^+$. When that message is forwarded to any other service, the taint labels according to $\mathcal{L}^-$ are removed and the ones determined by $\mathcal{L}^+$ are added. When a message is split, all resulting messages have the same taint labels as the original one and when messages are merged, the resulting message is tainted with the union of all individual taint labels.

\subsection{Static Model-Checking of Data Flows}

Dynamic taint tracking at runtime is sound under the assumption that the attacker does not know the message route definition, i.e. the control flow, but it is not complete, in the sense that it will only detect actual data leaks as they occur and not guarantee that a message route is free of data leaks in general. However, as it is important for users to know if a route may be interrupted by a policy, we use static model-checking to verify routes against policies.

For this purpose, we compile routes into Prolog, i.e. the same logic representation as policies. In Prolog, a route is represented as a directed acyclic graph, where each node represents one statement and edges refer to transitions between statements. Predicate \texttt{stmt} defines a statement and \texttt{succ(A,B)} defines \texttt{B} as a successor of \texttt{A}. This way the example route from \autoref{fig:example-route} can be written as the following (simplified) Prolog program:

\begin{minipage}{.45\columnwidth}
\begin{lstlisting}
stmt(sensor).
stmt(split).
stmt(log).
stmt(merge).
stmt(aggr).
stmt(mqueue).
\end{lstlisting}
\end{minipage}
\begin{minipage}{.45\columnwidth}
\begin{lstlisting}[firstnumber=7]
succ(sensor,split).
succ(split,log).
succ(split,merge).
succ(merge,aggr).
succ(log,aggr).
succ(aggr,mqueue).
\end{lstlisting}
\end{minipage}

Policies are likewise compiled into Prolog and determine valid and invalid flows in terms of allowed and forbidden labels entering services. Message routes can then be checked against policies by respectively exploring all paths in the graph which violate the policy. Each solution to the query is one counterexample of a possible data flow in a message route which does not comply with the policy. Listing \ref{lst:proof} shows the output displayed when a route violates a data flow policy.

\begin{lstlisting}[caption=Proof of a message route violating a policy,label=lst:proof,escapechar=!,otherkeywords={{Sensor_Messaging},dontPublishRaw,raw,{Outbound_Queue}},deletekeywords=is]
Route Sensor_Messaging is invalid because
service Outbound_Queue may receive label(s) [raw].
This is forbidden by rule dontPublishRaw

Example flows violating policy follow:
|-- sensor creates message labeled [raw]
|-- split receives message labeled [raw]
|-- log receives message labeled [raw]
|-- aggr receives message labeled [raw]
|-- mqueue receives message labeled [raw]
|-- fail!

\end{lstlisting}


\section{The LUCON Policy Language}

So far, we described how LUCON controls data flows in terms of abstract models, which provide the formal foundation of our policies. To be of any practical use, the framework must allow to write policies in a language that is easy to understand and supports the user in writing correct policies.

The LUCON policy language is a domain specific language (DSL) which serves two purposes: first, it comprises the actual data flow control rules, determining valid and invalid data flows and possibly binding them to obligations.
Second, it defines labels and describes services in terms of properties, capabilities and their taint propagation logic $\mathcal{L}^-$ and $\mathcal{L}^+$.

We define a grammar for the LUCON DSL in Eclipse XText. XText is a language creation framework that automatically creates lexer and parser from a context-free LL(*) grammar, along with IDE editors with syntax highlighting, auto-completion, and error checking.

The following is a simplified version of the LUCON DSL grammar. It defines the main concepts \emph{service} and \emph{rule}, which represent the service-specific taint propagation and the actual data flow rules. The effect of a rule is represented by a \emph{decision} that determines whether a message may be passed on or must be dropped. Optionally, a decision can be bound to an \emph{obligation}. Obligations are actions which must be executed successfully before the actual decision is enforced. If the execution of an obligation fails or the respective obligation is not supported by the system, the alternative decision stated by \texttt{otherwise} is taken.


\[
  \begin{array}{rcl}
  policy    &:=& \emph{rule}* \alt \emph{service}* \\
  service    &:=& \texttt{service \{} \\
  ~           & ~ &  ~~~~\texttt{id}~\emph{atom}\\
  ~           & ~ &  ~~~~\texttt{endpoint}~\emph{url}\\
  ~           & ~ &  ~~~~(\texttt{properties}~\emph{term}+)?\\
  ~           & ~ &  ~~~~(\texttt{capabilities}~\emph{term}+)? \texttt{~\}}\\
  rule  &:=& \texttt{flow\_rule \{} \\
  ~           & ~ &  ~~~~\texttt{when}~\emph{s}~\texttt{receives}~\emph{atom}\\
  ~           & ~ &  ~~~~\texttt{decide} ~\emph{decision} ~\texttt{\}}\\
  \emph{term} &:=& \text{Prolog term}\\
  \emph{atom} &:=& \text{Prolog atom}\\
  \emph{url} &:=& \text{Endpoint URL of a service}\\
  \emph{decision} &:=& \emph{effect}~(\emph{obligation})*\\
  \emph{effect} &:=& \texttt{allow} \alt \texttt{drop} \alt \texttt{error}\\
  \emph{obligation} &:=& \texttt{require} ~\emph{term} ~(\texttt{otherwise} ~\emph{term})?\\
  \emph{s} &:=& \text{Reference to a}~\emph{service}\\
  \end{array}
\]

\begin{figure}[tb]
	\centering
	\includegraphics[width=\columnwidth]{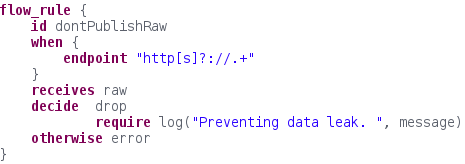}
	\caption{LUCON DSL rule in Eclipse IDE}
	\label{fig:rule-xtext}
  \vspace*{-.5cm}
\end{figure}


\autoref{fig:rule-xtext} shows a policy from the example scenario in section \ref{sec:introduction}. It includes an inline definition of the services it refers to -- in this example simply all services with an http(s) endpoint. The message label \texttt{raw} is stated as an atom (i.e., a 0-ary predicate) and marks raw sensor data. If that label has not been removed along the message route by some service that merges or blinds raw data records, the rule is triggered and will drop the message before it enters the respective endpoint. Before, the event will be logged, whereas \texttt{log} refers to a Java function which can be called from within the policy decision point and \texttt{message} refers to a predefined variable holding the content of the message. In case the execution of the obligation fails, the rule's effect is \texttt{error} which exceptionally terminates the message route.

LUCON policies are compiled into Prolog programs using the Xtend code generation framework, so that users only deal with the high level DSL, while the enforcement engine operates on the formal representation in Prolog. The representation of a policy in a logic model further allows reasoning over the policy itself to detect conflicting or incomplete rules and provides the basis for the aforementioned static model checking of message routes against data flow policies. Listing \ref{lst:prolog} shows the Prolog representation of the rule from \autoref{fig:rule-xtext}.

\begin{lstlisting}[caption=Prolog representation of a rule,label=lst:prolog]
regex(A,B,C) :- class("j.u.r.Pattern")
                    <- matches(A,B) returns C.
rule(dontPublishRaw).
has_target(dontPublishRaw, service15058189).
service(service15058189).
has_endpoint(service15058189,"http[s]?://.+").
receives_label(dontPublishRaw,raw).
has_decision(dontPublishRaw, dec).
has_effect(dec, drop).
has_obligation(dec,
      log("Preventing data leak. ", message)).
\end{lstlisting}

\section{Prototype Evaluation}


We implemented and evaluated a prototype of the LUCON policy framework to assess its application under real-world conditions.

\subsection{Implementation}

As a platform for our implementation we chose the \emph{Trusted IoT Connector} platform\footnote{\url{https://github.com/industrial-data-space/trusted-connector}} -- an open source platform based on the Karaf OSGi framework\footnote{\url{https://karaf.apache.org/}} that uses the Apache Camel message routing and mediation engine to forward messages between sensors and ''applications'' in form of Linux containers. While the Trusted Connector has been chosen for its security features, our implementation does not depend on it but would also be compatible with any other message router like Apache NiFi or Spring Integrations and other edge platforms like Eclipse Kura\footnote{\url{http://www.eclipse.org/kura/}}.

Apache Camel is a rule-based engine to route messages in the form of so-called \emph{Exchange} objects according to Enterprise Integration Patterns (EIP). Due to its support for more than 240 protocol adapters, including HTTP, OPC-UA, MQTT, it is well-suited for IoT scenarios where data from different sources must be unified. We hook into the Camel engine by implementing an \emph{interceptor} component that is called between each step in a message route and may drop, forward or alter any Exchange object. The interceptor acts as the Policy Enforcement Point (PEP) and interacts with the other components of the LUCON framework which have been implemented as OSGi services. If the message is allowed to pass, the PEP simply puts it back into the processing engine. If the decision is to drop the message, the interceptor removes it from the message route and in case of an error, it exceptionally terminates the route, allowing a graceful exception handling. Any obligation that is possibly bound to the policy decision refers to OSGi services that the PEP will invoke. As OSGi services are dynamic and can spawn and terminate at any time, the set of supported obligations may vary at runtime and policy authors must consider that the execution of obligations may fail by stating an alternative effect in the \texttt{otherwise} element.

The Policy Decision Point (PDP) includes a tuProlog engine to load policies as Prolog theories and run queries against them. tuProlog is a Java-native lightweight Prolog implementation that has been chosen because of its small footprint of only 294 KB and especially because of its ability to map Prolog predicates to Java functions. An example of such mappings is shown in line 1 of Listing \ref{lst:prolog} where a Prolog predicate \emph{regex} is defined by a call to the respective Java \emph{regex} function to support querying for regular expressions, e.g. over service endpoint URLs.


In total, the size of the LUCON policy engine amounts to a 3.1 MB OSGi bundle that is loaded into the Karaf platform, automatically detects all Camel instances and hooks its interceptor into their message routes. The policy parser and code generator is not part of that engine in order to keep its footprint low. This means, policy authors will write policies in LUCON DSL in a separate IDE and load the compiled policies into the engine. The LUCON IDE has been implemented as an Eclipse ''product'' i.e. a standalone version of the Eclipse IDE that includes the code generator and various assistants for authoring the policy.

\subsection{Data Flow Awareness of Services}

The most prevalent question for the integration of a policy framework is to which extend the existing IoT system must be aware of the framework and actively support it. LUCON requires only a single integration point, which is a hook into the message routing engine -- realized as a Camel interceptor in our prototype. In addition LUCON does not require services that are able to handle message labels. We distinguish services by three classes of message handling capabilities: \emph{agnostic}, \emph{preserving}, and \emph{active}.

\paragraph{Agnostic Services}
Agnostic services are unaware of any data flow control mechanism. That is, when messages are sent into an agnostic service, all message labels will be lost and the data flow tracing will break. Most existing services will fall into this category.\\
Agnostic services are supported by LUCON's capability to state transformation functions as part of the policy. As long as transformation functions are specified for a service, both runtime enforcement and static validation of routes will work as described above.

\paragraph{Preserving}
Even if they are not aware of any data flow control mechanism, some services are able to preserve labels attached to messages. That is, when data is sent into the service and retrieved at a later time, previously attached labels will still be intact and data flow tracing is not interrupted. As example for such services are databases or file systems which persist message labels along with data records.\\
As long as preserving services do not perform any operation that would change labels, no transformation function needs to be stated in the policy. Data flow tracing will not break at runtime and labels will be transported across service calls. Also static route validation will work, as the service does not affect labels and thus remains irrelevant with respect to path explorations in the message route graph.

\paragraph{Data Flow Aware Services}

Data flow aware services are able to actively modify message labels. While today, the vast majority of services is not data flow aware, an example of such services has been proposed in \cite{schuette2016}.\\
These services can modify message labels in a more complex way than could be expressed by transformation functions in the policy. The service in \cite{schuette2016} for instance, modifies message labels according to an internal ''taint logic'' that cannot be written as a transformation function. As a consequence, static route validation with data flow aware services is only possible if the service's labeling semantics is available in the same logic representation as the policy.

In general, data flow awareness of services directly relates to the trust in that service. A non-aware service does not require a high level of trust, since it would not be able to alter labels in a malicious way. Data flow aware services, on the contrary, are able to modify labels and could interfere with data flow that way. Consequently, data-flow aware services require additional mechanisms for trust establishment, such as a remote attestation or certification.

\subsection{Performance Evaluation}

\begin{figure}[tb]
  \centering
  \includegraphics[]{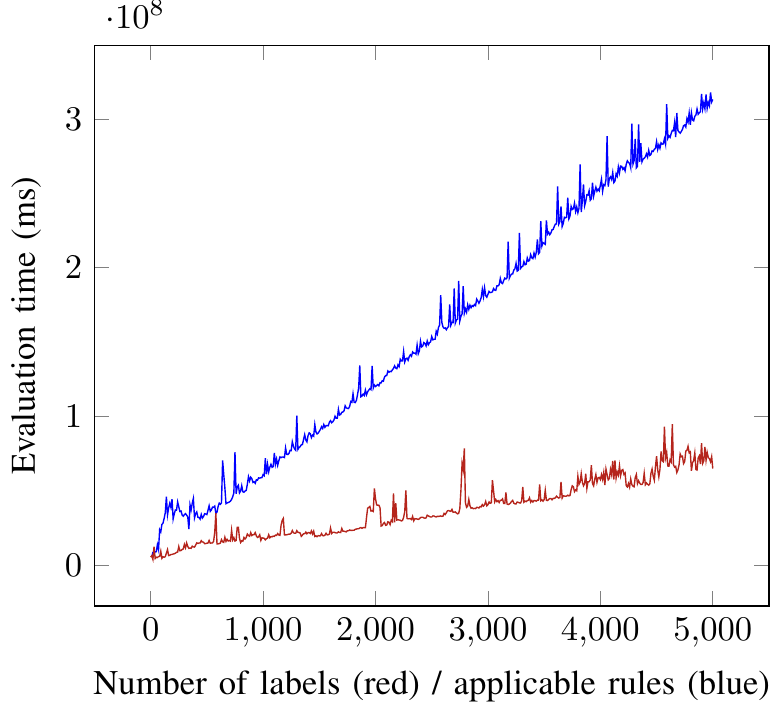}
  \caption{Policy decision time, scaling with rules (blue) and labels (red)}
  \label{fig:runtime-scaling-rules}
  \vspace*{-.44cm}
\end{figure}

The most critical metric for a policy engine is the time needed to evaluate a policy decision request. As each step in a message route requires a policy decision, the engine must not introduce unacceptable delays and must scale with an increasing number of services and rules. We evaluated how the runtime of the policy decision point for evaluating a decision request scales with the number of policies and services. While we consider a few dozens of rules to be a realistic size in most applications, we chose a test range of 1-5,000 rules and services. All rules were set up to match all services so that every decision request would require an evaluation of every single rule, which is the worst case. The tests were run against our prototype implementation which uses the Java-based tuProlog engine and does not include any runtime optimizations. As \autoref{fig:runtime-scaling-rules} shows, the time for evaluating the decision requests scales linearly with the number of rules within the analyzed range. For typical policy sizes of a few hundred rules, the evaluation takes approx. 12-15 ms. For a policy of 1,000 rules, it is still clearly below 50 ms and then increases linearly up to 150-200 ms for 5,000 rules. The red line in \autoref{fig:runtime-scaling-rules} shows how runtime scales with an increasing number of message labels and a constant number of 50 rules. As can be seen, the decision time only depends on the number of rules, but does not increase with more labels.

\begin{figure}[tb]
  \centering
  \includegraphics[]{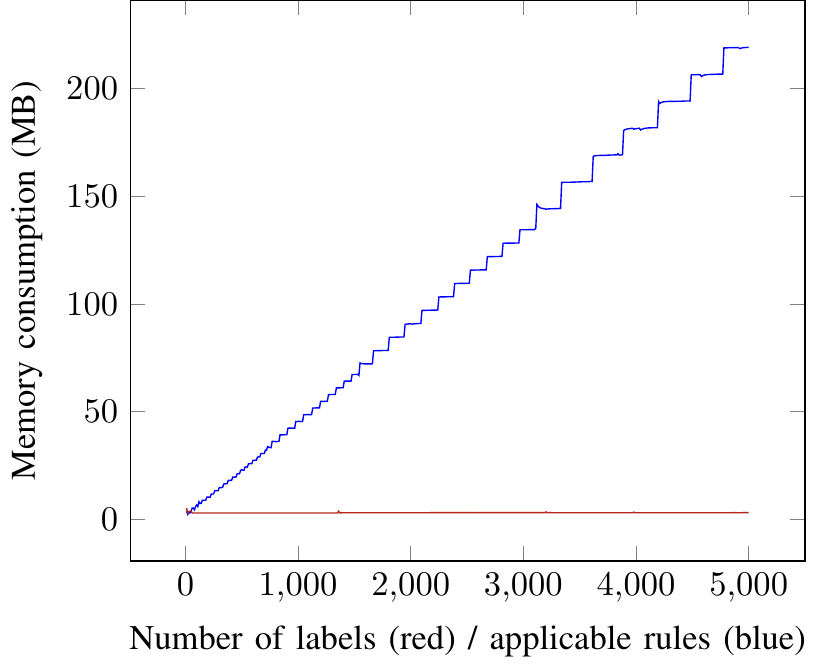}
  \caption{Memory for evaluating a policy decision. 1-5,00 rules/services/labels}
  \label{fig:memory-scaling-rules}
  \vspace*{-.44cm}
\end{figure}

The second metric of our performance evaluation is memory consumption. Here, we are especially interested if the framework is suited to run on typical IoT gateway devices or if the Prolog-based implementation it too memory-intensive for such applications. \autoref{fig:memory-scaling-rules} shows the memory consumption of the LUCON engine during a policy decision. Again, the blue line illustrates how memory consumption scales with an increasing number of rules and the red line indicates behavior with an increasing number of labels.

As expected, memory consumption scales linearly with the number of rules and constant with the number of labels, just as computation time does. The absolute numbers show that evaluating a policy of 50 rules requires less than 100 KB, while very large policies with thousands of rules may occupy several hundreds of megabyte of heap.

\section{Conclusions}
\label{sec:conclusions}

In this paper we introduced LUCON, a policy framework for controlling data flows in distributed message-based systems. LUCON extends the concept of usage control by the notion of data flows. In contrast to traditional information flow control frameworks which enforce a single security model or information classification scheme, our approach labels messages and monitors their usage in a taint analysis style, addressing an attacker model in which information leaks via side channels such as observation of the control flow are negligible. An automated formal verification of LUCON policies against message routes informs users upfront about possible policy violations and thus supports policy authors in writing correct rules. Proofs created by the formal verification support system audits, as they assert that message routes will not violate security and privacy requirements.

Our prototype shows that the approach of compiling policies and message routes into the same logic representation is both suitable for runtime enforcement and static verification, without the need to convert back and forth between different representations and possible semantic gaps. A major question was if the performance of a Prolog-based  evaluation engine can keep up with the demands of real-life systems with considerable high message throughput. Although performance impact of our prototype is notable, the measured delays in the range of 12-15 ms per policy decision are still in the range of typical network latency and suggest that with appropriate optimizations, the policy framework will easily be able to handle real world use cases.





\section*{Acknowledgement}

This work as been funded by the Federal Ministry for Economic Affairs and Energy (BMWi) in the project CAR-BITS (01MD16004B).

\bibliography{references}

\begin{thebibliography}{10}

\bibitem{Basin2012}
D.~Basin, M.~Harvan, F.~Klaedtke, and E.~Z\u{a}linescu.
\newblock Monpoly: Monitoring usage-control policies.
\newblock In {\em Proc. of the Second International Conference on Runtime
  Verification}, RV'11, pages 360--364, Berlin, Heidelberg, 2012.
  Springer-Verlag.

\bibitem{Basin2010}
D.~Basin, F.~Klaedtke, and S.~M\"{u}ller.
\newblock Policy monitoring in first-order temporal logic.
\newblock In {\em Computer Aided Verification}, volume 6174 of {\em Lecture
  Notes in Computer Science}, pages 1--18. Springer Berlin Heidelberg, 2010.

\bibitem{BellLaPadula73}
D.~E. Bell and L.~J. LaPadula.
\newblock Secure computer systems: Mathematical foundations.
\newblock MITRE Corporation, 1973.

\bibitem{Biba1977}
K.~J. Biba.
\newblock Integrity considerations for secure computer systems.
\newblock Technical report, MITRE Corp., 04 1977.

\bibitem{Chinis2013}
G.~Chinis, P.~Pratikakis, S.~Ioannidis, and E.~Athanasopoulos.
\newblock Practical information flow for legacy web applications.
\newblock In {\em Proc. of the 8th Workshop on Implementation, Compilation,
  Optimization of Object-Oriented Languages, Programs and Systems}, pages
  17--28. ACM, 2013.

\bibitem{Davis2010}
B.~Davis and H.~Chen.
\newblock Dbtaint: cross-application information flow tracking via databases.
\newblock {\em Proc. of WebApps}, 10, 2010.

\bibitem{Denning74}
D.~Denning.
\newblock A lattice model of secure information flow.
\newblock {\em Communications of the ACM}, 19(5):236--242, 1976.

\bibitem{Elrakaiby2014}
Y.~Elrakaiby and J.~Pang.
\newblock Dynamic analysis of usage control policies.
\newblock In {\em 11th Int. Conf. on Security and Cryptography (SECRYPT)},
  pages 88--100, Vienna, Austria, Nov. 2014.

\bibitem{Harvan2009}
M.~Harvan and A.~Pretschner.
\newblock State-based usage control enforcement with data flow tracking using
  system call interposition.
\newblock In {\em Network and System Security, 2009. NSS '09. Third
  International Conference on}, pages 373--380, Oct 2009.

\bibitem{hilty2007}
M.~Hilty, A.~Pretschner, D.~Basin, C.~Schaefer, and T.~Walter.
\newblock A policy language for distributed usage control.
\newblock In {\em ESORICS}, volume 4734, pages 531--546. Springer, 2007.

\bibitem{Hohpe2003}
G.~Hohpe and B.~Woolf.
\newblock {\em Enterprise Integration Patterns: Designing, Building, and
  Deploying Messaging Solutions}.
\newblock Addison-Wesley Longman Publishing Co., Inc., Boston, MA, USA, 2003.

\bibitem{katt2008}
B.~Katt, X.~Zhang, R.~Breu, M.~Hafner, and J.-P. Seifert.
\newblock A general obligation model and continuity: enhanced policy
  enforcement engine for usage control.
\newblock In {\em Proc. of the 13th ACM Symposium on Access Control Models and
  Technologies (SACMAT)}, pages 123--132. ACM, 2008.

\bibitem{Lazouski2010}
A.~Lazouski, F.~Martinelli, and P.~Mori.
\newblock Usage control in computer security: A survey.
\newblock {\em Computer Science Review}, 4(2):81 -- 99, 2010.

\bibitem{Myers1997}
A.~C. Myers and B.~Liskov.
\newblock A decentralized model for information flow control.
\newblock In {\em Proc. of the Sixteenth ACM Symposium on Operating Systems
  Principles}, SOSP '97, pages 129--142, New York, NY, USA, 1997. ACM.

\bibitem{Jif2001}
A.~C. Myers, L.~Zheng, S.~Zdancewic, S.~Chong, , and N.~Nystrom.
\newblock Jif: Java information flow.
\newblock Software release, July 2001].

\bibitem{Park2004}
J.~Park and R.~Sandhu.
\newblock The {$UCON_{ABC}$} usage control model.
\newblock {\em ACM Trans. Inf. Syst. Secur.}, 7(1):128--174, Feb. 2004.

\bibitem{Pasquier2016}
T.~Pasquier, J.~Bacon, J.~Singh, and D.~Eyers.
\newblock Data-centric access control for cloud computing.
\newblock In {\em Proc. of the 21st ACM on Symposium on Access Control Models
  and Technologies}, SACMAT '16, pages 81--88, New York, NY, USA, 2016. ACM.

\bibitem{Pasquier2015}
T.~F.~J. Pasquier, J.~Singh, D.~M. Eyers, and J.~Bacon.
\newblock Camflow: Managed data-sharing for cloud services.
\newblock {\em CoRR}, abs/1506.04391, 2015.

\bibitem{Pasquier2014}
T.~F. J.-M. Pasquier, J.~Bacon, and D.~Eyers.
\newblock {FlowK: Information Flow Control for the Cloud}.
\newblock {\em 6th Int. Conference on Cloud Computing Technology and Science
  (CloudCom)}, pages 1--8, 2014.

\bibitem{Pretschner2009a}
A.~Pretschner, M.~B{\"u}chler, M.~Harvan, C.~Schaefer, and T.~Walter.
\newblock Usage control enforcement with data flow tracking for x11.
\newblock In {\em Proc. of 5th Intl. Workshop on Security and Trust
  Management}, pages 124--137, 2009.

\bibitem{pretschner2006distributed}
A.~Pretschner, M.~Hilty, and D.~Basin.
\newblock Distributed usage control.
\newblock {\em Communications of the ACM}, 49(9):39--44, 2006.

\bibitem{Pretschner2009}
A.~Pretschner, J.~Ruesch, C.~Schaefer, and T.~Walter.
\newblock Formal analyses of usage control policies.
\newblock In {\em Availability, Reliability and Security, 2009. ARES '09},
  pages 98--105, March 2009.

\bibitem{Sabelfeld2010}
A.~Sabelfeld and A.~Russo.
\newblock {\em From Dynamic to Static and Back: Riding the Roller Coaster of
  Information-Flow Control Research}, pages 352--365.
\newblock Springer, Berlin, Heidelberg, 2010.

\bibitem{sandhu2003usage}
R.~Sandhu and J.~Park.
\newblock {Usage control: A Vision for Next Generation Access Control}.
\newblock In {\em MMM-ACNS}, volume 2776, pages 17--31. Springer, 2003.

\bibitem{schuette2016}
J.~Sch\"{u}tte and G.~S. Brost.
\newblock A data usage control system using dynamic taint tracking.
\newblock In {\em Proc. of the Int. Conference on Advanced Information Network
  and Applications (AINA), year=2016, month = mar}.

\bibitem{FlowCaml2003}
V.~Simonet.
\newblock The flow caml system.
\newblock Software release, July 2003.

\bibitem{Zhang2005}
X.~Zhang, F.~Parisi-Presicce, R.~Sandhu, and J.~Park.
\newblock Formal model and policy specification of usage control.
\newblock {\em ACM Transactions on Information and System Security (TISSEC)},
  (4), Nov. 2005.

\bibitem{Zhang2008}
X.~Zhang, J.-P. Seifert, and R.~Sandhu.
\newblock Security enforcement model for distributed usage control.
\newblock In {\em Sensor Networks, Ubiquitous and Trustworthy Computing
  (SUTC)}, pages 10--18, 2008.

\end{thebibliography}

\clearpage
\onecolumn
\section*{Appendix}

\begin{figure}[h]
\caption{Operational semantics of dynamically taint-controlled message routes}
\label{fig:operational-semantics}
\setlength{\jot}{1.2em}
\begin{gather*}
\infr[from]
  {\Delta, \cdot \vdash \texttt{get\_from(url)} \Downarrow \mu_m ~~~~~ \tau'[m \leftarrow \mathcal{L}^+(url)]~~~~~\iota'=\Sigma[pc+1]}
  {\langle \tau, \Sigma, \cdot, \Delta, \texttt{pc}, \iota\rangle, \texttt{from(url)} \to \langle \tau', \Sigma, \mu_m, \Delta, \texttt{pc+1}, \iota'\rangle}
\\
\infr[to]
  {\Delta, \mu_m \vdash \texttt{get\_from(url)} \Downarrow \mu_m ~~~~~~ \tau'[m \leftarrow \tau[m] \setminus \mathcal{L}^-(url) \cup \mathcal{L}^+(url)] ~~~~~ \iota'=\Sigma[pc+1]}
  {\langle \tau, \Sigma, \mu_m, \Delta, \texttt{pc}, \iota\rangle, \texttt{to(url)} \to \langle \tau', \Sigma, \mu_m, \Delta, \texttt{pc+1}, \iota'\rangle}
\\
\infr[choice (True)]
  {\Delta, \mu_m \vdash e \Downarrow 1 ~~~~~~ \Delta, \mu_m \vdash e_0 \Downarrow v_0 ~~~~~~ \iota'=\Sigma[v_0]}
  {\langle \tau, \Sigma, \mu_m, \Delta, \texttt{pc}, \iota\rangle, \mathsf{when}~e~\mathsf{then~goto}~e_0~\mathsf{otherwise~goto}~e_1 \to \langle \tau, \Sigma, \mu_m, \Delta, v_0, \iota'\rangle}
\\
\infr[choice (False)]
  {\Delta, \mu_m \vdash e \Downarrow 0 ~~~~~~ \Delta, \mu_m \vdash e_1 \Downarrow v_1 ~~~~~~ \iota'=\Sigma[v_1]}
  {\langle \tau, \Sigma, \mu_m, \Delta, \texttt{pc}, \iota\rangle, \mathsf{when}~e~\mathsf{then~goto}~e_0~\mathsf{otherwise~goto}~e_1 \to \langle \tau, \Sigma, \mu_m, \Delta, v_1, \iota'\rangle}
\\
\infr[split]
  {\Delta, \mu_m \vdash e \Downarrow m_0,...,m_n ~~~~~~ \tau' = \tau \cup \bigcup_{0 \leq i \leq n}\tau'[i]=\tau[m] ~~~~~~ \iota'=\Sigma[pc+1]}
  {\langle \tau, \Sigma, \mu_m, \Delta, \texttt{pc}, \iota\rangle, \mathsf{split}~e \to \langle \tau', \Sigma, \mu_m, \Delta, pc+1, \iota'\rangle}
~\\
\infr[aggregate]
  {\Delta, \bigcup_{i}\mu_i, \vdash e \Downarrow m ~~~~~~ \tau'[m] = \bigcup_{0 \leq i \leq n}\tau[i] ~~~~~~ \iota'=\Sigma[pc+1]}
  {\langle \tau, \Sigma, \bigcup_{i}\mu_i, \Delta, \texttt{pc}, \iota\rangle, \mathsf{aggregate}~e \to \langle \tau', \Sigma, \mu_m, \Delta, pc+1, \iota'\rangle}
\\
\infr[set-msg-prop]
  {\Delta, \mu_m, \vdash e \Downarrow v ~~~~~~ \mu'_m[k \leftarrow v] ~~~~~~ \iota'=\Sigma[pc+1]}
  {\langle \tau, \Sigma, \bigcup_{i}\mu_i, \Delta, \texttt{pc}, \iota\rangle, \mathsf{set\mhyphen msg\mhyphen prop}(k,e) \to \langle \tau, \Sigma, \mu'_m, \Delta, pc+1, \iota'\rangle}
  \\
\infr[set-env-prop]
  {\Delta, \mu_m, \vdash e \Downarrow v ~~~~~~ \Delta'[k \leftarrow v] ~~~~~~ \iota'=\Sigma[pc+1]}
  {\langle \tau, \Sigma, \bigcup_{i}\mu_i, \Delta, \texttt{pc}, \iota\rangle, \mathsf{set\mhyphen env\mhyphen prop}(k,e) \to \langle \tau, \Sigma, \mu_m, \Delta', pc+1, \iota'\rangle}
  \\
\infr[bean]
  {\Delta, \mu_m \vdash \texttt{bean(b)} \Downarrow \mu_m ~~~~~~ \tau'[m \leftarrow \tau[m] \setminus \mathcal{L}^-(b) \cup \mathcal{L}^+(b)] ~~~~~ \iota'=\Sigma[pc+1]}
  {\langle \tau, \Sigma, \mu_m, \Delta, \texttt{pc}, \iota\rangle, \texttt{bean(b)} \to \langle \tau', \Sigma, \mu_m, \Delta, \texttt{pc+1}, \iota'\rangle}
\\
\end{gather*}
\end{figure}

\end{document}